\documentclass[twocolumn,showpacs,preprintnumbers,amsmath,amssymb]{revtex4}
\usepackage{psfrag}
\usepackage{graphicx}
\usepackage{dcolumn}
\usepackage{bm}

\begin{document}

\preprint{}

\title{Theory of the collapsing axisymmetric cavity
               }

\author{J. Eggers$^1$, M.A. Fontelos$^2$, D. Leppinen$^3$, J.H. Snoeijer$^1$}

\affiliation{
School of Mathematics, University of Bristol, University Walk, 
Bristol BS8 1TW, UK$^1$ \\
Departamento de Matem\'{a}ticas, Consejo Superior de Investigaciones 
Cient\'{\i}ficas, C/ Serrano 123, 28006 Madrid, Spain$^2$ \\
School of Mathematics, University of Birmingham, Edgbaston 
Birmingham B15 2TT, UK$^3$
           }

\date{\today}

\begin{abstract}
We investigate the collapse of an axisymmetric cavity
or bubble inside a fluid of small viscosity, like water.
Any effects of the gas inside the cavity as well as of the
fluid viscosity are neglected. Using a slender-body description, we
show that the minimum radius of the cavity scales like 
$h_0 \propto t'^{\alpha}$, where $t'$ is the time from collapse. 
The exponent $\alpha$ very slowly approaches a universal value 
according to $\alpha=1/2 + 1/(4\sqrt{-\ln(t')})$. Thus, as 
observed in a number of recent experiments, the scaling can 
easily be interpreted as evidence of a single non-trivial scaling 
exponent. Our predictions are confirmed by numerical simulations.
\end{abstract}

\pacs{Valid PACS appear here}

\maketitle
Over the last decade, there has been considerable progress
in understanding the pinch-off of fluid drops, described 
by a set of universal scaling exponents, independent of the 
initial conditions \cite{E97,E05}. The driving is provided for by surface 
tension, the value of the exponents depend on the forces opposing 
it: inertia, viscosity, or combinations thereof. 
Bubble collapse appears to be a special case of an inviscid 
fluid drop breaking up inside another inviscid fluid, which
is a well studied problem \cite{CS97,DHL98,LL03}: the minimum 
drop radius scales like $h_0 \propto t'^{2/3}$, where $t'=t_0-t$
and $t_0$ is the pinch-off time. Thus, huge excitement 
was caused by the results of recent experiments on the pinch-off 
of an air bubble \cite{BWT05,GSRM05,TEK05,KMZN06,TEK06}, or the 
collapse of a cavity \cite{BMSSPL06} in water, which resulted in a 
radically different picture, in agreement with two earlier studies
\cite{LKL91,OP93}. As demonstrated in detail in \cite{TEK06}, the
air-water system corresponds to an inner ``fluid'' of vanishing inertia,
surrounded by an ideal fluid. 

\begin{figure}
\includegraphics[width=8cm]{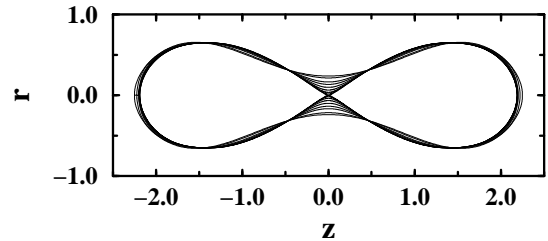}
\caption{Numerical simulation of the time evolution of bubble pinch-off from
initial conditions given by the shape with the largest waist. Pinch-off is 
initiated by surface tension, but the late stages are dominated 
by inertia, as observed experimentally \cite{TEK06}. 
\label{shape}   }
\end{figure}

Firstly, the scaling exponent $\alpha$ was found to be close to 1/2, 
(typical values reported in the literature are 0.56 \cite{KMZN06}
and 0.57 \cite{TEK06}), 
which means that breakup is much faster than in the fluid-fluid
case, and surface tension must become irrelevant as a driving force.
Secondly, the value of $\alpha$ appeared to depend subtly on the 
initial condition \cite{BMSSPL06}, and was typically found to be 
larger than 1/2. This raised the possibility of an ``anomalous'' 
exponent, selected by a mechanism as yet unknown. To illustrate the 
qualitative appearance of the pinch-off of a bubble, in Fig. \ref{shape}
we show a temporal sequence of profiles, using a full numerical simulation
of the inviscid flow equations \cite{LL03}. We confine ourselves 
to axisymmetric flow, which experimentally is found to be preserved 
down to a scale of a micron \cite{TEK06}, provided the experiment is aligned 
carefully \cite{KMZN06}. 

The only existing theoretical prediction \cite{GSRM05,BMSSPL06,GP06}
is based on treating the bubble as a (slightly perturbed)
cylinder \cite{LKL91,OP93}. This leads to the exponent being
1/2 with logarithmic corrections, a result which harks back 
to the 1940's \cite{L46}. Our numerics, to be reported below, 
are inconsistent with this result. Moreover, a cylinder is not
a particularly good description of the actual profiles 
(cf. Fig. \ref{shape}), as has been remarked before \cite{KMZN06}. 
In this Letter, we present a systematic expansion in the slenderness
of the cavity, which is found to lead to a self-consistent 
description of pinch-off. Our results are in excellent agreement 
with numerical simulations, and consistent with the experimentally
observed exponents. 

Our approach is based on the standard description \cite{AL65,S02} of 
slender cavities, an assumption that is tested self-consistently
by showing that the cavity's axial extension is greater than 
its radius. The inviscid, irrotational, incompressible 
flow ${\bf u}=\nabla \phi$
outside the cavity of length $2L$ is written as 
\begin{equation}
\phi = \int_{-L}^L\frac{C(\xi)d\xi}{\sqrt{(z-\xi)^2+r^2}},
\label{pot}
\end{equation}
where $C(\xi)$ is a line distribution of sources 
to be determined. The length $L$ will later 
drop out of the description of the pinch region, as indeed 
(\ref{pot}) is not expected to be good near the ends of the bubble. 
For a slender geometry, $\partial_z\phi\ll\partial_r\phi$, 
and the radial velocity, again using slenderness, is easily
evaluated to be $\partial_r\phi = -2C(z)/h(z)$. 

The equation of motion for the collapsing cavity of radius 
$h(z,t)$ is $\partial_t h \approx u_r$, and thus 
$\dot{a}(z,t)\approx -4C$, where $a = h^2$ and the dot denotes the time 
derivative. Finally, an 
equation of motion for $C$ comes from the Bernoulli equation, 
evaluated at the free surface \cite{LL84}. We then arrive at 
\begin{equation}
\int_{-L}^L\frac{\ddot{a}(\xi,t)d\xi}{\sqrt{(z-\xi)^2+a(z,t)}} = 
\frac{\dot{a}^2}{2a} + 4\Delta p /\rho, 
\label{bernoulli}
\end{equation}
where $\Delta p = \gamma\kappa + const$ is the pressure difference 
across the cavity \cite{note}. In the two-fluid problem, the surface tension 
$\gamma$, multiplied by the mean curvature $\kappa \approx 1/h$,
drives the problem. The capillary pressure will however turns out 
to be subdominant, so the last term in (\ref{bernoulli}) can 
effectively be neglected. Note that the resulting equation is 
invariant under a rescaling of both space and time, as both remaining terms
are inertial (describing acceleration and convection of a fluid element).
Thus dimensional arguments do not work, and a more detailed analysis 
is needed to fix the scaling exponent. Note that (\ref{bernoulli}) does not
conserve the volume of the cavity, whereas Fig. \ref{shape} assumes an 
incompressible gas inside the bubble. This however only affects the rounded 
ends of the bubble. 

Our aim is to explain the observed scaling behavior of the minimum cross 
section $a_0=a(0,t)$, as well as of the axial length scale 
$\Delta$ of the profile, which can be characterized by the inverse curvature
$\Delta\equiv (2a_0/a''_0)^{1/2}$, where $a''_0=a''(0,t)$ and 
the prime denotes denotes a derivative with respect to $z$. 
Experiments as well as our own simulations show that 
$a_0\approx A t'^{2\alpha}$ and $\Delta \approx D t'^{\beta}$
with $\beta < \alpha$, thus the radius is small compared to the 
axial extend at the minimum. This means that $a(0,t)$ can be neglected 
relative to $\xi^2\approx\Delta^2$ in the denominator of the integral, 
except 
near the position $\xi=0$ of the minimum. In other words, 
the integral is dominated by {\it local} contributions near the 
minimum. This will permit us to find equations of motion for the 
minimum in terms of local quantities alone.

As shown later, $\ddot{a}(\xi,t)$ goes to zero over the axial scale 
$\Delta$. Thus the integral at $z=0$ can be approximated as 
\[
\ddot{a}_0\int_{-\Delta}^{\Delta}\left[\xi^2+a_0\right]^{-1/2} d\xi 
\approx \ddot{a}_0\ln(2\Delta^2/a_0). 
\]
An arbitrary factor inside the logarithm depends on the exact shape
of $\ddot{a}(\xi,t)$; it can be determined empirically, but in fact
becomes subdominant in the limit $a''_0\rightarrow 0$. 
However, we now need another equation for the (time-dependent) 
width $\Delta$ to close the description. To that end we evaluate 
the second derivative of (\ref{bernoulli}) at $z=0$. 

The contribution of the left hand side of (\ref{bernoulli}) is
\[
\int_{-\Delta}^{\Delta}\ddot{a}(\xi,t)
\left[\frac{2\xi^2-a_0}{\sqrt{\xi^2+a_0}^5} - 
\frac{a''_0}{2\sqrt{\xi^2+a_0}^3}\right]d\xi.
\]
For a slender profile, $a''_0$ is subdominant, but the 
integral over the first term in angular brackets conspires to 
give zero in the limit $a_0\rightarrow 0$, so the second
term has to be considered as well, and $\ddot{a}(\xi,t)$ has to 
be expanded beyond the constant term:
$\ddot{a}(\xi,t)=\ddot{a}_0 + \ddot{a''_0}\xi^2/2$. Thus using 
the same reasoning as before, and keeping in mind that $a'_0=0$,
we find for the second derivative of the integral
\begin{eqnarray}
\nonumber
\lefteqn{\int_{-\Delta}^{\Delta}
\left[\frac{(\ddot{a}_0+\ddot{a}''_0\xi^2/2)
(2\xi^2-a_0)}{\sqrt{\xi^2+a_0}^5} - \right. } \\
&& \left. \frac{\ddot{a}_0a''_0}{2\sqrt{\xi^2+a_0}^3}\right]d\xi \approx
 \left[\ddot{a}''_0\ln\left(\frac{4\Delta^2}{e^3a_0}\right) 
-2\frac{\ddot{a}_0a''_0}{a_0}\right]. \nonumber
\end{eqnarray}

Equating this with the second derivative of the right hand side of 
(\ref{bernoulli}), $(\dot{a}^2/(2a))''$, which is readily 
computed in terms of $a_0$ and $\Delta$, yields the desired 
second equation. It is slightly more convenient to rewrite the 
results as equations for the time-dependent exponents 
\begin{equation}
2\alpha\equiv -\partial_{\tau}a_0/a_0, \quad 
2\delta \equiv -\partial_{\tau}a''_0/a''_0,
\label{exp}
\end{equation}
where $\tau \equiv -\ln t'$ and $\beta = \alpha-\delta$.
Note that (\ref{exp}) is going to be the ``true'' definition 
of the (time-dependent) exponents, which agrees with a local 
power-law fit. The result is 
\begin{eqnarray}
\label{motion1}
&&\left(\alpha_{\tau}+\alpha-2\alpha^2\right)\ln(\Gamma_1/a''_0) 
= -\alpha^2, \\
\label{motion2}
&&\left(\delta_{\tau}+\delta-2\delta^2\right)\ln(\Gamma_2/a''_0) =
2\alpha-3\alpha^2-2\alpha\delta+2\alpha_{\tau}, \quad
\end{eqnarray}
where the subscript denotes the $\tau$-derivative. 

The scaling factors $\Gamma_1,\Gamma_2$ have to be determined 
empirically, but only make a subdominant contribution as 
$a''_0$ goes to zero. The time dependence of $a''_0$ is 
best found from integrating 
\begin{equation}
\ln(a''_0)_{\tau}=-2\delta.
\label{lnmotion}
\end{equation}
An analysis of (\ref{motion1})-(\ref{lnmotion}) shows that the approach to 
the singularity corresponds to an {\it unstable} fixed point as
$\tau\rightarrow\infty$. As usual, this is the result of the 
freedom in the choice of singularity time $t_0$, see for 
example \cite{LL03}. The limit $\alpha=1/2$ thus has to be imposed
onto the system in order to find the physically relevant solution.
From the first equation, one finds that $\alpha$ approaches 1/2
from above, while the second equation says that $\delta$ goes to zero, 
but remains positive. This guarantees the self-consistency of our 
approximation, although $\beta$ approaches $\alpha$ in the limit. 
However, the approach of $\alpha$ and $\beta$ toward their limiting 
values is exceedingly slow, as seen from the expansion
\begin{equation}
\alpha=1/2+\frac{1}{4\sqrt{\tau}}+\frac{\Gamma}{\tau}, \quad
\delta =\frac{1}{4\sqrt{\tau}} + O(\tau^{-3/2}),
\label{asymp}
\end{equation}
where $\Gamma$ is a constant which reflects the arbitrariness of the timescale
in (\ref{bernoulli}). Thus the value of $\Gamma$ necessarily depends on 
initial conditions. However to leading order $\alpha$ approaches its limiting 
value in a universal fashion. Finally, for the self-consistency of our 
analysis we need that the dimensionless parameter $a''_0$ goes to zero toward 
pinch-off, as is indeed found from (\ref{lnmotion}), owing to the slowness
with which $\delta$ converges toward zero. 

We now turn to a detailed comparison with full numerical simulations, 
not relying on any slenderness assumption, by focusing on the late stages 
of the pinch-off event shown in Fig. \ref{shape}.  To this end a suitably 
modified version of the boundary integral code developed to examine 
inviscid droplet pinch-off \cite{LL03} was used, as originally 
reported in \cite{LLE05}. This involved two important modifications:
First, the boundary value operator (cf. Equation (11) in \cite{LL03})
has a zero eigenvalue in the case of the absence of an inner fluid, 
corresponding to a change in the bubble volume. This singularity is
analytically removed before the boundary integral operator is inverted,
fixing the bubble volume. Second, due to the rapidity of bubble pinch-off, 
the adaptive time-stepping used for droplet pinch-off in \cite{LL03} 
was replaced by a time-step halving procedure with error estimation.

\begin{figure}
\psfrag{t'}{\Large$t'$}
\includegraphics[width=7cm]{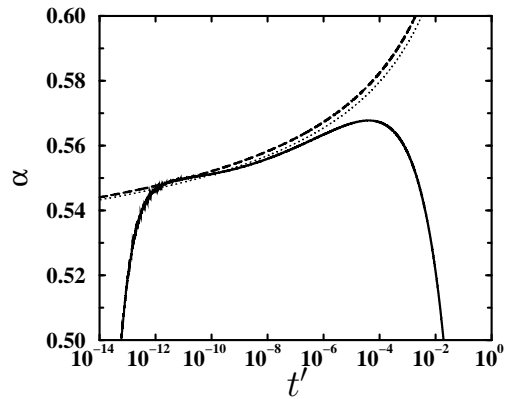}
\caption{A comparison of the exponent $\alpha$ between full 
numerical simulations of bubble pinch-off (solid line) and the 
leading order asymptotic theory with $\Gamma=0$ (dashed line) and the 
second order correction with $\Gamma=0.1$ (dotted line).
\label{compare}   }
\end{figure}

A comparison of the numerical simulations with (\ref{asymp}) is given 
in Fig. \ref{compare}. Using equation (\ref{exp}), the value of 
$\alpha$ from the numerical simulations can be calculated as 
$\alpha = t' \partial_{t'} h_0/h_0$, and the pinch-off time $t_0$ is 
estimated from the numerical data. The solid curve in Fig. \ref{compare} 
is the data from the numerical simulation, the dashed curve is the 
leading order prediction given by equation (\ref{asymp}) with $\Gamma=0$, 
and the dotted curve includes the adjustable constant with $\Gamma=0.1$.

Data from the numerical simulations can be divided into three regimes. 
From approximately $10^{-12} < t' < 10^{-4}$ the bubble is
considered to be in the asymptotic regime, and it is seen that there 
is very good agreement between the numerical data and the asymptotic 
theory: the leading order theory with $\Gamma=0$ accurately predicts the 
extremely slow decrease in the numerically determined value of 
$\alpha$, and the second order correction with $\Gamma=0.1$ improves the
agreement between the asymptotic theory and the numerical data. 
Equally good agreement was found for numerical runs using other
initial conditions, provided that $\Gamma$ was adjusted, as it is
expected to depend on initial conditions, as also observed 
in experiment \cite{BMSSPL06}.
Time $t'> 10^{-4}$ corresponds to a transitional regime where the 
bubble adjusts from an initial state where surface tension
is required to initiate pinch-off, to an asymptotic state where 
surface tension is irrelevant. Time $t' \sim 10^{-12}$ represents the 
threshold of the numerical simulations: extremely large interfacial
velocities acting over ever-decreasing lengthscales, ultimately puts 
a limit on the validity of the numerical simulations.

Gordillo et al. \cite{GSRM05,GP06} have previously predicted that 
the minimum bubble radius $h_0$ should scale with $t'$ according 
to $t' \propto h_0^{2}\sqrt{-\ln h_0^{2}}$, using a method that in 
many respects is similar to ours \cite{GP06}. However, the crucial 
difference is that they do not treat the axial length scale $\Delta$ 
as a dynamical variable as we do, but effectively identify $\Delta$ with
some outer length scale. Indeed, if one replaces $a''_0$ by $a_0$ in
(\ref{motion1}), one recovers the scaling result of \cite{GP06}.  
\begin{figure}
\psfrag{a}{\Large$\ddot{a}$}
\psfrag{z}{\Large$z$}
\includegraphics[width=7cm]{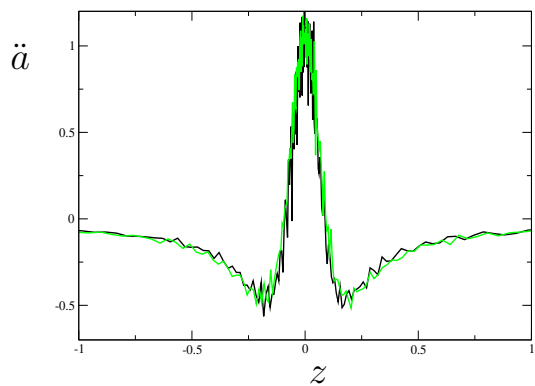}
\caption{A normalized graph of $\ddot{a}=\partial^2 h^2(z,t)/\partial t^2$ 
as given by the full numerical simulations, for two different initial 
conditions, and at $t'=3.8\times 10^{-10}$ (black line) and 
$t'=2.1\times 10^{-10}$ (green line). 
\label{peak} }
\end{figure}
The conceptual difference between the two approaches is illustrated 
further by Fig. \ref{peak}, which shows the central peak of $\ddot{a}$
from the full numerical simulation. The value of $\ddot{a}$ rapidly 
drops to zero, effectively providing the cutoff of the integral 
(\ref{bernoulli}) at an axial length $\Delta$, which is shrinking 
like $t'^{\beta}$. So far, we have not been able to identify the 
logarithmic corrections of $\beta$ in our full numerical simulations, 
since computing the axial scale is much more demanding than computing $h_0$. 

\begin{figure}
\psfrag{t'}{\Large$t'$}
\includegraphics[width=7cm]{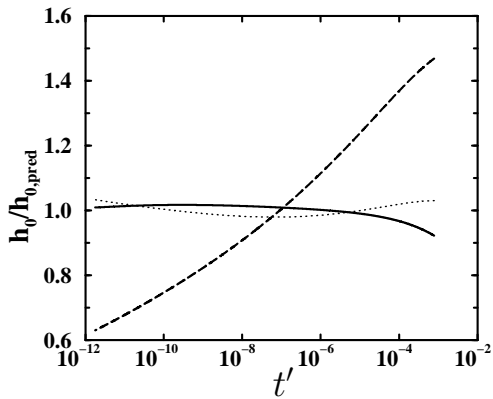}
\caption{A normalized graph of $h_0/h_{0,pred}$ where
$h_{0,pred}$ is predicted according to the theory presented by 
Gordillo et al. \cite{GSRM05} (dashed line),
a least square approximation \cite{LLE05} (dotted line), 
and the current asymptotic theory with
$\alpha = 1/2 + \frac{1}{4 \sqrt{\tau}}$ (solid line).
\label{compare2}   }
\end{figure}

In Fig. \ref{compare2} we plotted the numerically computed 
minimum radius $h_0$, divided by the universal part of the 
present theory (full line), and that of \cite{GP06} (dashed line). 
If normalized by an appropriate constant, the result should be unity. 
Namely, (\ref{asymp}) with $\Gamma=0$ is equivalent to 
$h_{0,pred} \propto t'^{1/2} \sqrt{e^{-\sqrt{-\ln t'}}}$, while 
the theory in \cite{GSRM05} amounts to 
$h_{0,pred} \propto t'^{1/2}/(-\ln h_0^{2})^{1/4}$. While the present
theory agrees extremely well with numerics without the use of any
adjustable constant, the theory in \cite{GP06} varies by approximately 
$\pm 50\%$ over the range of $t'$ plotted.

In our earlier numerical simulations \cite{LLE05}, as well as in 
most experimental papers \cite{BWT05,KMZN06,TEK06}, the data 
for the minimum radius was represented by adjusting a single 
exponent $\bar{\alpha}$. Although Fig. \ref{compare} clearly shows
that the exponent is slowly varying, this subtle feature is 
difficult to detect in a more conventional plot like Fig. \ref{compare2}. 
To demonstrate this point, we have determined an effective exponent
$\bar{\alpha}=0.559$ from a least-square fit to the numerical data,
a value which is close to those observed experimentally 
\cite{KMZN06,TEK06}. 
In essence, $\bar{\alpha}$ can be viewed as the average over $\alpha$ 
values shown in Fig. \ref{compare}. The resulting fit (dotted line)
gives a surprisingly good description of the data, as a result of the 
extremely slow variation of $\alpha$. It also highlights the need
for more sophisticated plots like Fig. \ref{compare} in the 
interpretation of future (experimental) data.

To summarize, we have developed an asymptotic theory for the 
collapse of an axisymmetric cavity. A novel feature of this theory
is a slow variation of the scaling exponents, whose leading order
contributions are universal. The slowness of the approach explains
the experimental observation of apparently new scaling exponents, 
whose value may depend weakly on initial conditions.
It remains to calculate the entire
form of the central peak of $\ddot{a}$, which according to 
Fig. \ref{peak} is universal. This will determine the values
of the constants $\Gamma_1$ and $\Gamma_2$.
Other challenges are the inclusion of non-axisymmetry \cite{KMZN06} 
and viscosity \cite{TEK06} into the theoretical description.

\acknowledgments
We thank J. Lister for his continued support, valuable insight,
and very detailed comments on the manuscript, as well as 
J.M. Gordillo for discussions. S. Thoroddsen made his experiments
available to us prior to publication, for which we are grateful. 
JHS acknowledges financial support from a Marie Curie European
Fellowship FP6 (MEIF-CT2006-025104).


\begin{thebibliography}{0}
\bibitem{note} Equation (\ref{bernoulli}), as it stands, 
is however ill-posed. 
Namely, if $a(z,t)$ in the denominator of the integral operator
is assumed constant, the operator can be inverted by Fourier
transform. One finds that the high-wavenumber modes of $\ddot{a}$ 
grow like $\exp(k\sqrt{a})$, so the evolution (\ref{bernoulli})
will soon be polluted by short-wavelength noise. However, this 
problem can be dealt with completely by adding a small ``damping''
term $\epsilon \ddot{a}''a$ to the right hand side. Choosing 
$\epsilon = 10^{-3}$, we find that the evolution is completely 
stable, yet the damping term and its derivatives is always uniformly
smaller than the other terms by more than 4 orders of magnitude.
\end{thebibliography}
\end{document}